\documentclass[aps,prx,twocolumn, superscriptaddress,amsmath,amssymb,longbibliography]{revtex4-2}
\usepackage{xcolor}
\usepackage{amsmath}
\usepackage{graphicx}% Include figure files
\usepackage{dcolumn}% Align table columns on decimal point
\definecolor{webblue}{HTML}{1111aa}
\usepackage[
colorlinks=true,
linkcolor=webblue,
citecolor=webblue,
urlcolor=webblue,
]{hyperref}
\usepackage{bm}% bold math
\usepackage[T1]{fontenc}
\usepackage{hyphenat}

\begin{document}

\preprint{APS/123-QED}

\title{Observation of spin-free interatomic orbital angular momentum in a chiral crystal}

\author{Dongjin Oh}
\email[Corresponding Author:$~$]{djeeoh@snu.ac.kr}
\affiliation{Department of Physics and Astronomy, Seoul National University, Seoul, 08826, Korea}

\author{Sungsoo Hahn}
\affiliation{MAX IV Laboratory, Lund University, 22100 Lund, Sweden}

\author{Chiara Pacella}
\affiliation{Department of Physics and Astronomy, University of Bologna, Bologna, Italy}
\affiliation{Max Planck Institute for the Structure and Dynamics of Matter, Hamburg, Germany}

\author{Junseo Yoo}
\affiliation{Department of Physics and Astronomy, Seoul National University, Seoul, 08826, Korea}

\author{Angel Rubio}
\affiliation{Max Planck Institute for the Structure and Dynamics of Matter, Hamburg, Germany}
\affiliation{Nano-Bio Spectroscopy Group, Departmento de Física de Materiales, Universidad del País Vasco, San Sebastián, Spain}
\affiliation{Initiative for Computational Catalysis (ICC) and Center for Computational Quantum Physics (CCQ), The Flatiron Institute, New York, NY, USA}

\author{Domenico Di Sante}
\email[Corresponding Author:$~$]{domenico.disante@unibo.it}
\affiliation{Department of Physics and Astronomy, University of Bologna, Bologna, Italy}

\author{Changyoung Kim}
\email[Corresponding Author:$~$]{changyoung@snu.ac.kr}
\affiliation{Department of Physics and Astronomy, Seoul National University, Seoul, 08826, Korea}

\begin{abstract}
The inherent spin-orbit interaction of electrons inevitably couples spin to the orbital angular momentum (OAM), posing a fundamental challenge to spin-free orbital transport. Here, we propose a novel strategy to achieve spin-decoupled OAM states in crystalline solids. Using angle-resolved photoemission spectroscopy (ARPES), we resolve well-isolated $s$-orbital bands in a chiral Te crystal, clearly separated from the $p$-orbital manifold. Combined circular dichroism ARPES and first-principles calculations reveal that these bands host OAM arising exclusively from interatomic hopping, with no intra-atomic contribution. Spin-resolved ARPES further confirms the absence of spin angular momentum (SAM), providing decisive evidence of spin-free OAM states. These findings establish the existence of OAM without spin polarization in crystalline solids and highlight the essential role of interatomic OAM. This work provides a general framework for designing spinless OAM states, opening an opportunity toward pure orbital currents for orbitronics.
%By combining circular-dichroism and spin-resolved angle-resolved photoemission spectroscopy with first-principles calculations, we demonstrate that $s$-orbital electrons in a chiral Te crystal acquire finite OAM through interatomic hopping while exhibiting no spin polarization. These results provide direct evidence for OAM states without spin involvement in crystalline solids and underscore the essential role of interatomic OAM. Our work establishes a general framework for engineering spin-free OAM states in solids, opening an opportunity toward pure orbital currents for spin-free orbitronics.

\end{abstract}

\maketitle

Orbital angular momentum (OAM, $\vec{L}$) and spin angular momentum (SAM, $\vec{S}$) constitute fundamental quantum degrees of freedom of electrons that govern the magnetic properties of condensed matter systems \cite{hirst_microscopic_1997}. While the contribution of OAM has long been regarded as negligible due to its quenching in crystalline environments \cite{kittel_introduction_2005}, its pivotal role in emergent correlated and topological phenomena has recently gained increasing recognition in quantum material research \cite{choi_observation_2023,el_hamdi_observation_2023,liu_orbital_2021,redekop_direct_2024,han_orbital_2023,unzelmann_orbital-driven_2020,hagiwara_orbital_2025,fukami_challenges_2025,oh_p-wave_2026}. %For example, orbital transport phenomena---conceived as the orbital counterparts of spin transport---have been experimentally realized \cite{choi_observation_2023,el_hamdi_observation_2023}. Moreover, orbital magnetization driven by OAM has been shown to underpin a range of emergent quantum phases, including those characterized by the quantum anomalous Hall effect, the fractional quantum Hall effect, and orbital multiferroicity in moiré systems and rhombohedral graphene \cite{liu_orbital_2021,redekop_direct_2024,han_orbital_2023}. 
These advances underscore that, in stark contrast to the conventional notion of orbital quenching, OAM should be treated as an indispensable degree of freedom in modern condensed matter physics.

\begin{figure}[htbp!]
	\includegraphics[width=8.1cm]{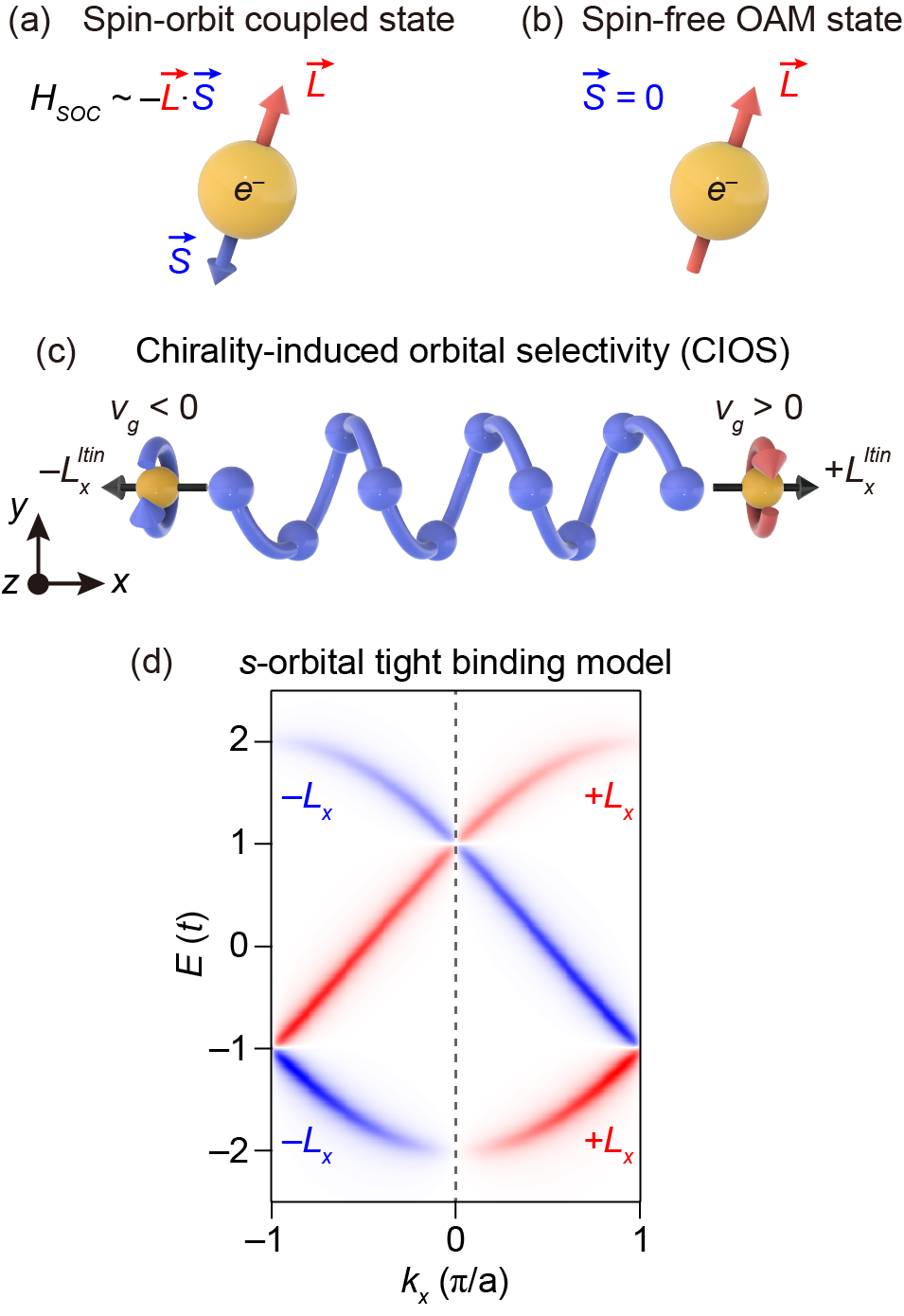}
	\caption{  
        (a,b) Schematics of spin-orbit-coupled (a) and spin-free orbital angular momentum (OAM) states (b). Red, blue arrows and yellow spheres indicate OAM, spin angular momentum (SAM), and electrons, respectively. 
        (c) Illustration of chirality-induced orbital selectivity (CIOS) in a right-handed chiral chain. Blue and red circular arrows denote the $x$-component of $\vec{L}^{Itin}$.
        (d) Tight-binding band structure of $s$-orbital electronic states in a single helical chiral chain. OAM values are color coded.
        }
        \label{fig:1}
\end{figure}

While OAM and SAM originate from fundamentally distinct origins, disentangling their respective contributions is generally nontrivial. In the presence of atomic spin-orbit coupling (SOC) $H_{SOC}\sim-\vec{L} \cdot \vec{S}$, the OAM and SAM of an ensemble of electrons become intrinsically entangled [Fig.~\ref{fig:1}(a)]. Consequently, under finite SOC, the emergence of OAM is inherently accompanied by a corresponding spin component. For this reason, materials composed of light elements with small atomic number $Z$, and thus weak SOC, are widely regarded as promising platforms for maximizing (minimizing) orbital (spin) contributions \cite{jo_gigantic_2018,go_orbitronics_2021}. As such, establishing a fundamental framework for suppressing SAM contributions is a central challenge in the realization of genuinely spin-free OAM states [Fig.~\ref{fig:1}(b)].

A refined classification of OAM provides key insights into the realization of spinless OAM states. Electronic OAM can be categorized into two distinct subtypes \cite{thonhauser_orbital_2005,burgos_atencia_orbital_2024}. The first is atomic OAM ($\vec{L}^{Atom}$), which originates from the OAM of atomic orbitals. Accordingly, $s$-orbital electrons are typically considered to carry no $\vec{L}^{Atom}$. For other orbitals, $\vec{L}^{Atom}$ inevitably couples to SAM via atomic SOC, posing a fundamental obstacle to spin-decoupled OAM states. The second contribution is itinerant OAM ($\vec{L}^{Itin}$), arising from the geometric phase of Bloch wavefunctions acquired through interatomic electron hopping \cite{thonhauser_orbital_2005,burgos_atencia_orbital_2024}. Because this contribution does not depend on the atomic orbital character, even $s$-orbital electrons can exhibit a finite $\vec{L}^{Itin}$ \cite{busch_orbital_2023,burgos_atencia_orbital_2024}. Moreover, since atomic orbitals with $s$ character carry no $\vec{L}^{Atom}$, they are not directly coupled to atomic SOC and do not inherently lead to spin polarization.
Taken together, these points indicate that SOC-free $s$-orbital electrons may sustain finite $\vec{L}^{Itin}$ in the absence of $\vec{L}^{Atom}$, offering a viable pathway toward spinless OAM states. However, despite increasing theoretical attention \cite{bhowal_orbital_2021,pezo_orbital_2022,pezo_orbital_2023,cysne_orbital_2022,busch_orbital_2023}, experimental verification of this itinerant component remains scarce, as prior studies have predominantly emphasized the role of $\vec{L}^{Atom}$ \cite{park_orbital-angular-momentum_2011,park_chiral_2012,park_orbital_2012,kim_microscopic_2013,go_intrinsic_2018,unzelmann_momentum-space_2021,choi_observation_2023}.

To realize $\vec{L}^{Itin}$ states without spin polarization under this framework, we focus on chiral materials featuring helical lattice geometry, where interatomic electron hopping can naturally generate $\vec{L}^{Itin}$. In this Letter, we investigate the $s$-orbital states in a chiral Te crystal. A combination of angle-resolved photoemission spectroscopy (ARPES)-based experiments and first-principles calculations reveals that these states carry finite OAM of purely interatomic origin while exhibiting no SAM. Our results not only underscore the critical role of $\vec{L}^{Itin}$ in crystalline solids but also point to a compelling route to engineering spin-free OAM states by leveraging this interatomic contribution.

%Using angle-resolved photoemission spectroscopy (ARPES), we identify well-isolated $s$-orbital bands that are clearly separated from the $p$-orbital manifolds. Combining circular dichroism ARPES (CD-ARPES) with first-principles calculations, we further demonstrate that these bands exclusively carry $\vec{L}^{Itin}$, with no contribution from $\vec{L}^{Atom}$. Moreover, complementary spin-resolved ARPES (SARPES) measurements unambiguously confirm the absence of SAM, providing decisive evidence of a spin-free OAM texture. Our findings underscore the essential role of $\vec{L}^{Itin}$ in condensed matter systems and establish a compelling strategy for engineering electronics states with spinless OAM. By opening an opportunity toward pure orbital currents devoid of spin contributions, this work represents a significant milestone in the pursuit of spin-free orbitronics.

\begin{figure*}[htbp!]
	\includegraphics[width=16.5cm]{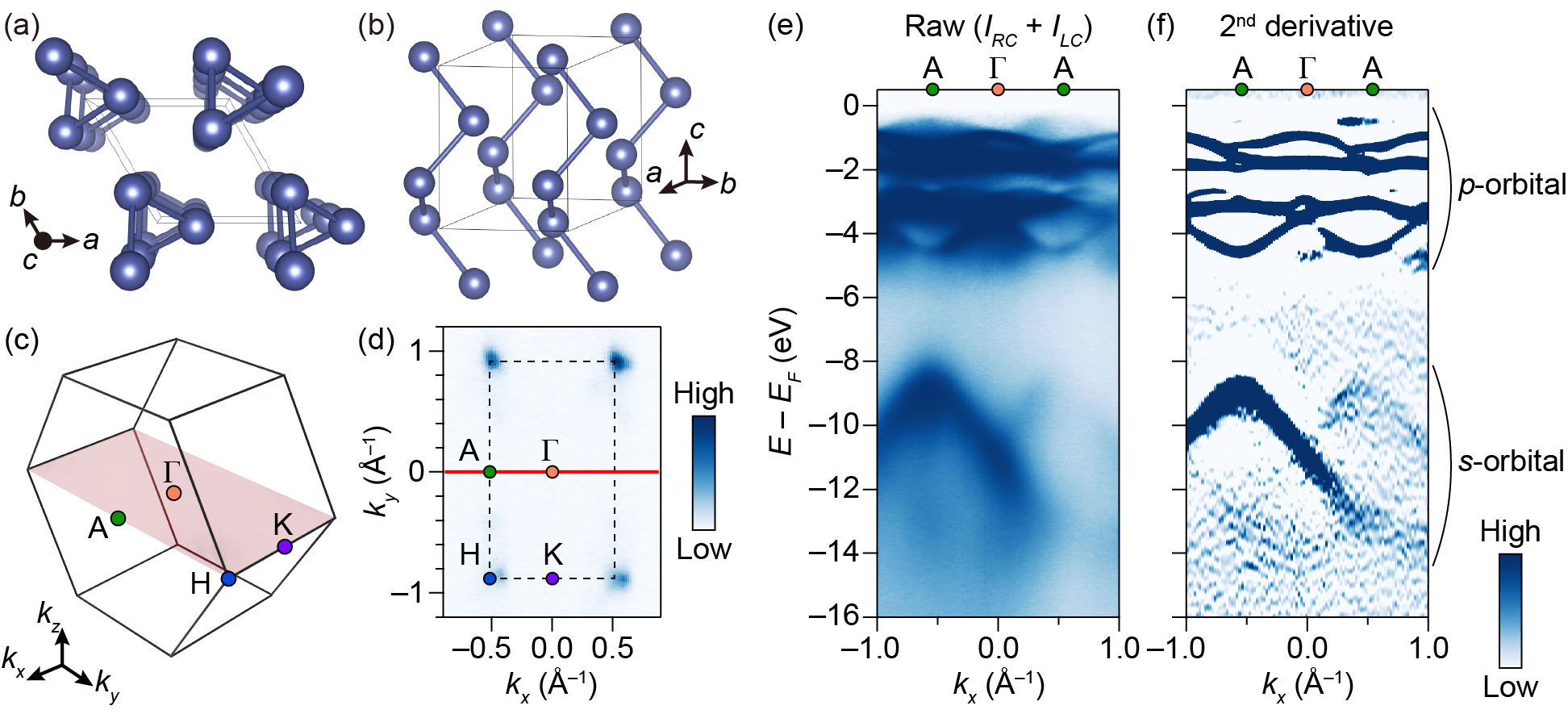}
	\caption{  
		(a,b) Top (a) and side (b) views of the crystal structure of elemental Te. 
        (c) Brillouin zone of Te. The red plane indicates the momentum regions accessible by ARPES at 90 eV photon energy. The $k_{x}$, $k_{y}$, and $k_{z}$ are defined with respect to the coordinate of the ARPES experimental geometry.
        (d) Fermi surface of Te measured at a photon energy of 90 eV. The black dashed rectangle marks the boundary of the first Brillouin zone. The red solid line denotes the plane of incidence in the ARPES experimental geometry. 
        (e) Raw ARPES spectrum of Te along the A-$\Gamma$-A high-symmetry $k$-path. 
        (f) Corresponding second-derivative ARPES spectrum. The ARPES spectra shown in (d), (e), and (f) are obtained by averaging data acquired with right- and left-circularly polarized light. 
        }
        \label{fig:2}
\end{figure*}

As an ideal model system, we consider a chiral chain with a helical lattice structure, in which the microscopic mechanism underlying $\vec{L}^{Itin}$ can be understood intuitively [Fig.~\ref{fig:1}(c)]. In this chiral lattice, electrons propagating along the chain axis ($x$-axis) follow helical hopping path. Therefore, in a right handed chiral chain, electrons moving in the $+x$ ($-x$) direction with positive (negative) group velocity $v_{g}$ acquire $+L_{x}^{Itin}$ ($-L_{x}^{Itin}$) [Fig.~\ref{fig:1}(c)]. This microscopic mechanism underlying the formation of $\vec{L}^{Itin}$ is known as chirality-induced orbital selectivity (CIOS) \cite{gobel_chirality-induced_2025}, which can be regarded as the orbital analogue of chirality-induced spin selectivity \cite{naaman_chiral_2019}. This real space picture can be naturally translated into the band representation. For a chiral chain possessing threefold screw-rotation symmetry, a tight-binding model predicts three electronic bands, as shown in Fig.~\ref{fig:1}(d) \cite{gobel_chirality-induced_2025}. In this momentum space representation, the sign of $L_{x}^{Itin}$ follows the group velocity of the bands: bands with $\partial E/\partial k>0$ exhibit $+L^{Itin}_{x}$, whereas those with $\partial E/\partial k<0$ exhibit $-L^{Itin}_{x}$. This simple model thus provides clear and intuitive insight into how chiral electron hopping, dictated by the underlying lattice geometry, gives rise to $\vec{L}^{Itin}$. %Guided by this understanding, we turn our attention to a simple chiral material with a helical structure to explore $\vec{L}^{Itin}$.

In this work, we particularly focus on the $s$-orbital states in a helical lattice to verify the existence of $\vec{L}^{Itin}$. Because $s$-orbital electrons inherently carry no $\vec{L}^{Atom}$, this offers a clear advantage by enabling the elimination of atomic contribution and the direct probing of $\vec{L}^{Itin}$. We note that elemental Te crystal meets all the essential criteria for realizing our proposed framework. As illustrated in Fig.~\ref{fig:2}(a) and~\ref{fig:2}(b), Te crystalizes in a chiral chain structure composed of monoatomic helical lattices. %In addition, although the low-energy $5p$ bands of Te reside in the low-energy regime \cite{joannopoulos_electronic_1975,hirayama_weyl_2015}, a distinct set of three well-isolated $s$-orbital bands is theoretically anticipated at higher binding energies \cite{peng_elemental_2014}. 
However, since prior studies have concentrated on the low-energy $p$ states \cite{gatti_radial_2020,sakano_radial_2020,nakayama_band_2017,nakayama_observation_2024}, these $s$-orbital bands have so far remained experimentally unexplored. Therefore, the first objective of this work is the experimental identification of the $s$-orbital bands in chiral Te. 

To experimentally explore the $s$-orbital bands in Te, we performed ARPES experiments. Consistent with previous studies, Te crystal can be easily cleaved along the (10$\bar{1}$0) surface due to its van der Waals nature \cite{sakano_radial_2020,gatti_radial_2020,nakayama_band_2017,nakayama_observation_2024}. This cleavage plane enables access to the $\Gamma$-A-H-K high symmetry $k$-path, highlighted as the red plane in Fig.~\ref{fig:2}(c). Indeed, the measured Fermi surface using 90 eV photon energy shown in Fig.~\ref{fig:2}(d) agrees well with prior result \cite{gatti_radial_2020}, confirming the well cleaved (10$\bar{1}$0) surface.

Figure~\ref{fig:2}(e) displays the energy-momentum dispersion along the A-$\Gamma$-A high-symmetry $k$-path, corresponding to the direction parallel to the Te chiral chain axis. Interestingly, alongside the 5$p$ bands observed within the $E-E_{F}=-6$ to 0 eV window, we clearly resolve highly dispersive 5$s$ bands in the deeper binding energy range of $E-E_{F}=-15$ to $-8$ eV. The dispersion of the $s$-orbital states is more clearly visualized in the second-derivative spectra shown in Fig.~\ref{fig:2}(f). Remarkably, the experimentally observed dispersion of the $s$-orbital bands along the Te chain axis is in striking agreement with the tight-binding band structure of a single chiral chain [Fig.~\ref{fig:1}(d)]. This excellent agreement between experiment and the tight-binding model prediction establishes the $s$-orbital electronic states in Te crystal as an ideal and experimentally accessible platform for validating our proposed route toward spin-free OAM states.

\begin{figure*}[htbp!]
	\includegraphics[width=16.4cm]{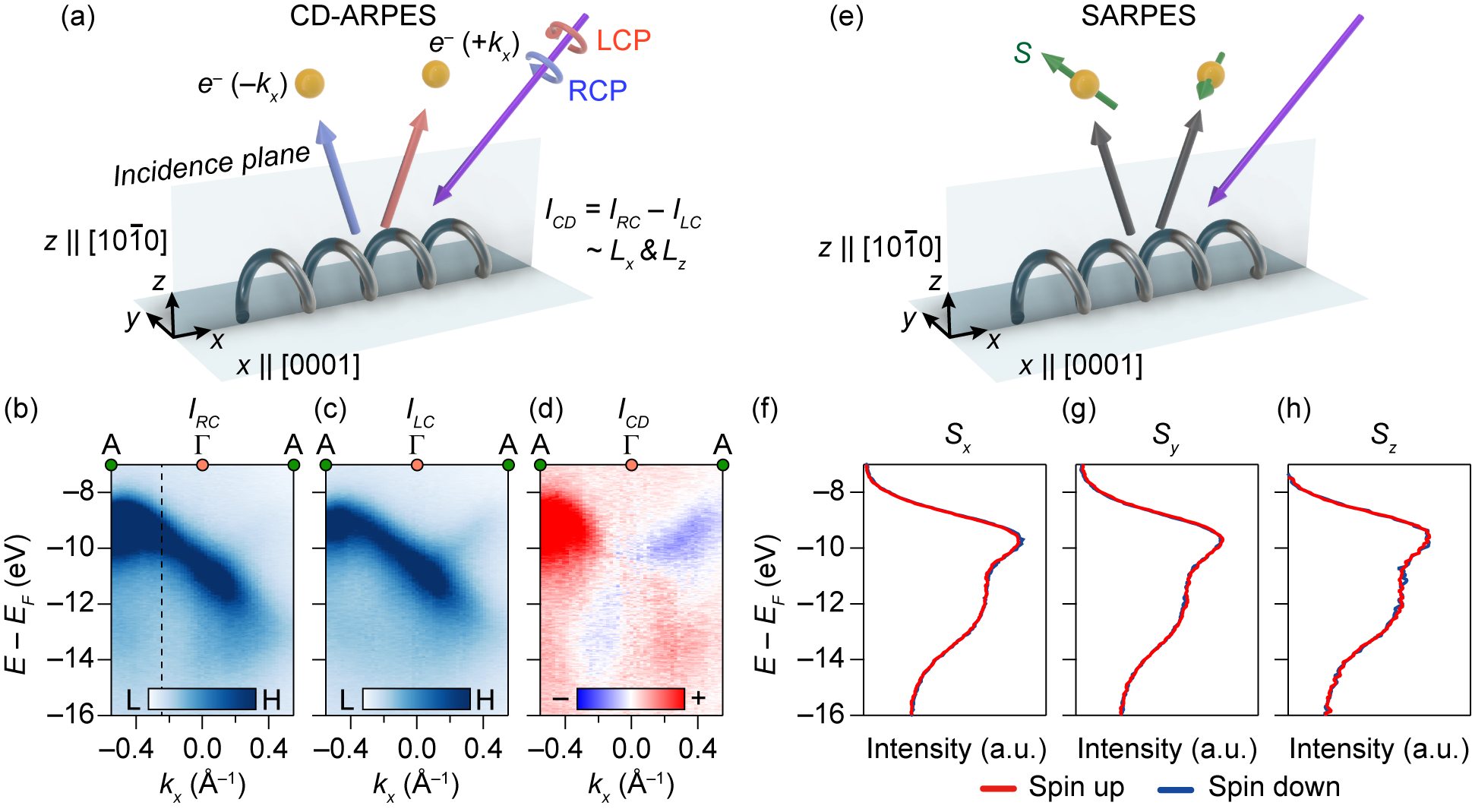}
	\caption{  
		(a) Schematic illustration of CD-ARPES geometry. The purple arrow denotes incident light with right- and left-circular polarization (RCP and LCP). 
        (b,c) ARPES spectra of the $s$-orbital bands measured using RCP (b) and LCP (c). 
        (d) Corresponding CD-ARPES spectrum ($I_{CD}=I_{RCP}-I_{LCP}$).
        (e) Schematic illustration of the SARPES geometry.
        (f-h) Spin-resolved energy distribution curves (SEDCs) for the $S_{x}$ (f), $S_{y}$ (g), and $S_{z}$ (h) components. The red and blue curves represent the spin-up and spin-down contributions, respectively. The SEDCs are extracted along the black dashed line indicated in (b). RCP light was used for the SARPES measurements.
        }
        \label{fig:3}
\end{figure*}

We further performed CD-ARPES experiments to verify whether the $s$-orbital bands of Te carry $\vec{L}^{Itin}$ as predicted by the tight-binding model. CD-ARPES technique can probe OAM components that are parallel to the light propagation direction \cite{moser_toy_2023,figgemeier_imaging_2025,oh_interplay_2025}. Therefore, in the typical ARPES geometry illustrated in Fig.~\ref{fig:3}(a), where the incident light impinges obliquely within the plane of incidence ($xz$-plane), CD-ARPES allows access to the $x$- and $z$-components of OAM ($L_{x}$ and $L_{z}$). Following this methodology, we intentionally align the [0001] axis of Te along the $x$-axis of our experimental coordinate [Fig.~\ref{fig:3}(a)] in order to probe OAM component parallel to the chain. Furthermore, we carefully aligned crystallographic axis such that the [10$\bar{1}$0] direction lies in the plane of incidence. Therefore, in this experimental coordinate, $L_{x}$ and $L_{z}$ are defined as the components parallel to $k_{x}$ and $k_{z}$, respectively [Fig.~\ref{fig:2}(c)].

Figure~\ref{fig:3}(b) shows the ARPES spectra measured using right-circular polarization (RCP). Under RCP illumination, only the $s$-orbital band with negative $v_{g}$ is observed. In contrast, as shown in Fig.~\ref{fig:3}(c), a band with positive $v_{g}$ becomes visible under left-circular polarization (LCP) light, while the photoelectron intensity of the negative $v_{g}$ band is reduced. These ARPES spectra acquired under RCP and LCP illumination give rise to the corresponding CD-ARPES spectrum, which clearly exhibits two bands with opposite CD signs crossing each other [Fig.~\ref{fig:3}(d)]. The magnitude of CD at $E-E_{F}=-9.2$ eV and $k_{x}=\pm 0.4$ $\text{\AA}^{-1}$ reaches $\mp 17$\%, indicating a sizable and physically meaningful CD signal (Supplemental Material). In contrast to the two sharply resolved crossing bands, the lowest-energy 5\textit{s} band exhibits substantially weaker and broader spectral weight. Consequently, its CD signal could not be clearly resolved in the present measurements.

\nocite{kresse1996efficient}
\nocite{kresse1996vasp}
\nocite{blochl1994paw}
\nocite{kresse1999paw}
\nocite{perdew1996pbe}
\nocite{hobbs2000noncollinear}
\nocite{steiner2016soc}
\nocite{jain2013materialsproject}
\nocite{marzari1997wannier}
\nocite{mostofi2008wannier90}
\nocite{pizzi2020wannier90}
\nocite{post_wan}
\nocite{Dongwook_anom_pos}

Importantly, the observed CD-ARPES spectrum, which maps the momentum distribution of CD along the A-$\Gamma$-A $k$-path, reflects the intrinsic CD signal. This high-symmetry CD-ARPES spectrum was obtained along the $k_{y}=0$ $\text{\AA}^{-1}$ line [red solid line in Fig.~\ref{fig:2}(d)], which is aligned with the plane of incidence in our experimental geometry [Fig.~\ref{fig:3}(a)]. In such CD-ARPES configuration, the extrinsic geometrical CD contribution vanishes at $k_{y}=0$ $\text{\AA}^{-1}$ \cite{PhysRevLett.132.196402}, whereas away from $k_{y}=0$ $\text{\AA}^{-1}$, the intrinsic CD can be hindered by the extrinsic geometrical CD. Therefore, the measured CD-ARPES spectrum presented in Fig.~\ref{fig:3}(d) represents the intrinsic CD signal of the $s$-orbital electronic states in Te crystal, which is proportional to $L_{x}$ and $L_{z}$.

We further conducted SARPES measurements to examine the SAM character of the $s$-orbital states. The SARPES measurements were conducted in the same experimental geometry as the CD-ARPES experiments [Fig.~\ref{fig:3}(e)], and spin-resolved energy distribution curves (SEDCs) were extracted at $k_{x}=-0.25$ $\text{\AA}^{-1}$, as indicated by the black dashed line in Fig.~\ref{fig:3}(b). Figures~\ref{fig:3}(f)-\ref{fig:3}(h) display the measured SEDCs projected onto the $S_{x}$, $S_{y}$, and $S_{z}$ components, respectively. Interestingly, in contrast to the presence of a finite CD signal, SEDCs for spin-up and spin-down contributions show no discernible difference for all three orthogonal spin components. These results indicate the absence of SAM polarization in the $s$-orbital states of Te, unlike the pronounced radial SAM texture observed in the low-energy $p$-orbital bands \cite{sakano_radial_2020,gatti_radial_2020}.

Our complementary CD-ARPES and SARPES measurements clearly reveal the emergence of an intrinsic CD signal in $s$-orbital states without any accompanying SAM polarization. It is also worth noting that the observed momentum distribution of CD shown in Fig.~\ref{fig:3}(d) is reminiscent of $\vec{L}^{Itin}$ character predicted by CIOS in a single chiral chain [Fig.~\ref{fig:1}(d)], capturing the band crossing between the $+L_{x}^{Itin}$ and $-L_{x}^{Itin}$ states. This suggest that the $s$-orbital electrons hosted by the helical lattice of the Te can indeed form OAM states induced by $\vec{L}^{Itin}$. 

\begin{figure}[t!]
	\includegraphics[width=8.5cm]{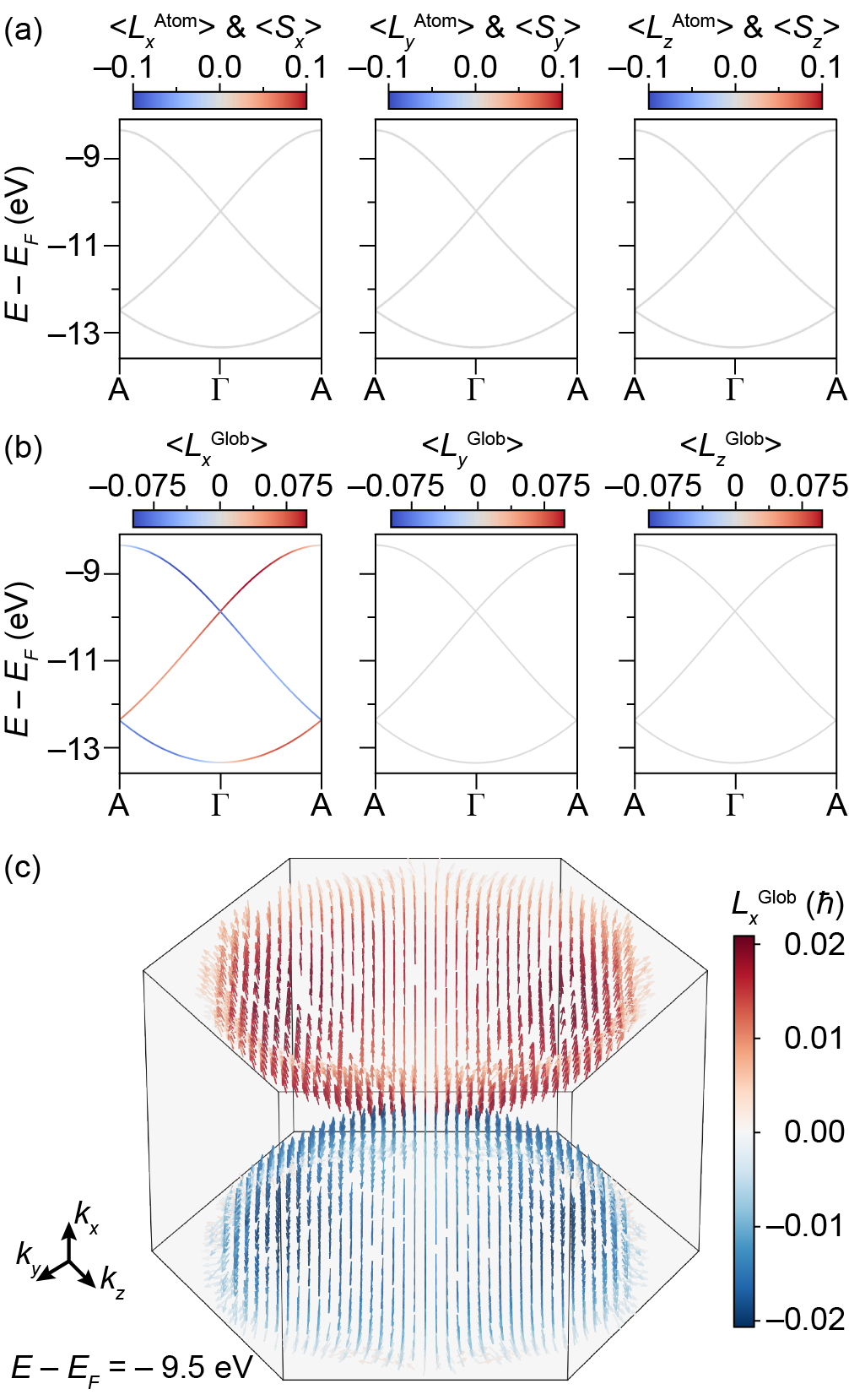}
	\caption{  
		OAM and SAM characteristics of $s$-orbital states in a right-handed Te crystal.
        (a) Expectation values of $L_{x}^{Atom}$ and $S_{x}$ (left), $L_{y}^{Atom}$ and $S_{y}$ (middle), and $L_{z}^{Atom}$ and $S_{z}$ (right). Each $\vec{L}^{Atom}$ is evaluated within ACA.
        (b) Expectation values of $L_{x}^{Glob}$ (left), $L_{y}^{Glob}$ (middle), and $L_{z}^{Glob}$ (right) calculated using modern theory of orbital magnetization.
        (c) Texture of $\vec{L}^{Glob}$ mapped onto the iso-energy surface at $E-E_{F}=-9.5$ eV. The color scale represents the magnitude of $L_{x}^{Glob}$.
        }
        \label{fig:4}
\end{figure}

Beyond comparison with $L_{x}^{Itin}$ texture of a single chiral chain model, we further investigate the OAM and SAM characteristics of the $s$-orbital states in the Te crystal using first-principles calculations, to elucidate how these properties evolve from an isolated chain to an array of chiral chains (Supplemental Material). We examine the OAM character of $s$-orbital states using two different theoretical frameworks: atomic-centered approximation (ACA) and modern theory of orbital magnetization \cite{thonhauser_orbital_2005,bhowal_orbital_2021,pezo_orbital_2022,pezo_orbital_2023,cysne_orbital_2022,busch_orbital_2023}. In the former approach, $\vec{L}^{Itin}$ is completely neglected, and the entire OAM contribution is approximated as $\vec{L}^{Atom}$. In contrast, the modern theory allows one to compute the global OAM ($\vec{L}^{Glob}$), in which both $\vec{L}^{Atom}$ and $\vec{L}^{Itin}$ are simultaneously present. Therefore, by comparing the OAM character calculated using these two different approaches, we can gain insight into how $\vec{L}^{Atom}$ and $\vec{L}^{Itin}$ incorporate to the $s$-orbital states.

Figure~\ref{fig:4}(a) shows the $\vec{L}^{Atom}$ character of $s$-orbital states obtained using ACA method as well as SAM polarization. As intuitively expected, these calculations further confirm that the $s$-orbital bands indeed do not carry $\vec{L}^{Atom}$. Moreover, these $s$-orbital bands do not exhibit spin splitting as these bands are not affected by atomic SOC and therefore remain spin-degenerate. Consequently, no spin polarization appears in these bands. These first-principles calculations are consistent with the SARPES results, which experimentally demonstrate the absence of spin polarization [Figs.~\ref{fig:3}(f)-\ref{fig:3}(h)].

Remarkably, whereas the $\vec{L}^{Atom}$ texture obtained using the ACA vanishes, the $\vec{L}^{Glob}$ calculated within the modern theory framework yields a finite value as shown in Fig.~\ref{fig:4}(b). It is worth noting that although $\vec{L}^{Glob}$ contains contributions from both $\vec{L}^{Atom}$ and $\vec{L}^{Itin}$, the $s$-orbital states considered here show no contribution from $\vec{L}^{Atom}$, as evidenced in Fig.~\ref{fig:4}(a). Thus, $\vec{L}^{Glob}$ can effectively be treated as $\vec{L}^{Itin}$ for $s$-orbital electrons. Indeed, $L_{x}^{Itin}$ texture calculated using tight-binding model for a single chiral chain [Fig.~\ref{fig:1}(d)] and $L_{x}^{Glob}$ from first-principles calculation for Te crystal [Fig.~\ref{fig:4}(b), left] show excellent agreements. In contrast to $L_{x}^{Glob}$, the $y$- and $z$-components of $\vec{L}^{Glob}$ ($L_{y}^{Glob}$ and $L_{z}^{Glob}$) vanish, as shown in Fig.~\ref{fig:4}(b), indicating that $\vec{L}^{Glob}$ is predominantly polarized along the Te chiral-chain axis. Consequently, the iso-energy surface at $E-E_{F}=-9.5$ eV exhibits a dipole-like $\vec{L}^{Glob}$ texture [Fig.~\ref{fig:4}(c)], with $\vec{L}^{Glob}$ pointing predominantly along the $+x$ direction in the $+k_{x}$ region and along the $-x$ direction in the $-k_{x}$ region.

%In contrast to the tight-binding model for a single chiral chain [Fig.~\ref{fig:1}(d)], first-principles calculations for chiral Te crystal reveal finite $y$- and $z$-components of $\vec{L}^{Glob}$ ($L_{y}^{Glob}$ and $L_{z}^{Glob}$), as shown in Fig.~\ref{fig:4}(b). This deviation likely originates from interchain electron hopping. Indeed, as shown in Fig.~\ref{fig:4}(c), the iso-energy surface at $E-E_{F}=-9.5$ eV exhibits pronounced three-dimensional dispersion across the Brillouin zone, directly evidencing electronic coupling between neighboring chains. Correspondingly, the $\vec{L}^{Glob}$ texture projected onto this iso-energy surface displays a complex structure, while preserving the sign reversal of $L_{x}^{Glob}$ between positive and negative $k_{x}$ regions. These results demonstrate that first-principles calculations faithfully capture the intricate $\vec{L}^{Itin}$ texture arising from three-dimensional electronic hybridization in bulk chiral Te.

It is worth noting that the close correspondence between our CD-ARPES measurements and first-principles calculations provides strong evidence for a pure $\vec{L}^{Itin}$ texture, without any $\vec{L}^{Atom}$ contribution. This constitutes the first spectroscopic evidence for the existence of $\vec{L}^{Itin}$ and highlights its fundamental importance in crystalline solids. Nevertheless, it should be noted that the CD-ARPES signal contains not only information on the initial-state OAM but also contributions arising from photoemission matrix-element effects, and its interpretation therefore requires appropriate caution. A systematic CD-ARPES simulation incorporating the relevant experimental geometry, photon energy, and final-state effects may provide a useful route toward disentangling these contributions and achieving a more quantitative understanding of the measured dichroism.
Furthermore, SARPES measurements demonstrate that engineering $s$-orbital electrons offers a promising route to realizing spin-decoupled OAM states governed by $\vec{L}^{Itin}$. Although the relevant $s$-orbital states in chiral Te crystal lie far from the Fermi level, limiting their immediate practical applicability, our findings demonstrate that harnessing interatomic OAM in \textit{s}-orbital-derived electronic states provides a promising route toward realizing spin-free OAM states and offers a new design principle for future orbitronic materials.

%%%%%%%%%%%%%

\textit{Acknowledgments}---We acknowledge the fruitful discussion with Dongwook Go, Junhee Shin, and Hosub Jin. This work was supported by the National Research Foundation of Korea (NRF) grant funded by the Korean government (MSIT) (No. 2022R1A3B1077234) and also by GRDC(Global Research Development Center) Cooperative Hub Program through the National Research Foundation of Korea(NRF) funded by the Ministry of Science and ICT(MSIT) (RS-2023-00258359). Support from the Institute of Applied Physics at Seoul National University is acknowledged. We acknowledge the MAX IV Laboratory for beamtime on the Bloch
beamline under proposal 20252331. Research conducted at MAX IV, a Swedish national user facility, is supported by Vetenskapsrådet (Swedish Research Council, VR) under contract 2018-07152, Vinnova (Swedish Governmental Agency for Innovation Systems) under contract 2018-04969 and Formas under contract 2019-02496. D.D.S acknowledges MUR funding within the FIS2 (n. 1236, 01-08-2023) Project no. FIS-2023-00144 (CUP J53C25001880001).This work was supported by the Cluster of Excellence "Advanced Imaging of Matter" (AIM), Grupos Consolidados y de Alto Rendimiento UPV/EHU, and Gobierno Vasco (IT1453-22). We acknowledge support from the Max Planck–New York City Center for Non-Equilibrium Quantum Phenomena. The Flatiron Institute is a division of the Simons Foundation.

D.O., S.H., and C.P. contributed equally. %\cite{kresse1996efficient,kresse1996vasp,blochl1994paw,kresse1999paw,perdew1996pbe,hobbs2000noncollinear,steiner2016soc,jain2013materialsproject,marzari1997wannier,mostofi2008wannier90,pizzi2020wannier90,post_wan,Dongwook_anom_pos}

\textit{Data availability}---The data that support the findings of this article are openly available at \cite{Dongjin_observation}.

\bibliography{references}

\end{document}